\documentclass{aa}

\usepackage{txfonts}
\usepackage{graphicx}

\title{Non-thermal radio emission from single hot stars}
\author{S. Van Loo \and M.C. Runacres \and R. Blomme}
\institute{Royal Observatory of Belgium, Ringlaan 3, B-1180 Brussel, Belgium}
\offprints{S. Van Loo, \email{Sven.VanLoo@oma.be}}

\date{Received  / Accepted} 

\abstract{
We present a theoretical model for the non-thermal radio emission from single
hot stars, in terms of synchrotron radiation from electrons accelerated in
wind-embedded shocks.  The model is described by five independent parameters
each with a straightforward physical interpretation. Applying the model to a
high-quality observation of \object{Cyg~OB2~No.~9} (O5 If), we obtain meaningful
constraints on most parameters. The most important result is that the outer
boundary of the synchrotron emission region must lie between 500 and 2200
stellar radii. This means that shocks must persist up to that distance.
We also find that relatively weak shocks (with a compression ratio
$< 3$) are needed to produce the observed radio spectrum. These results are
compatible with current hydrodynamical predictions. Most of our models
also show a relativistic electron fraction that increases outwards. This 
points to an increasing efficiency of the acceleration mechanism, perhaps 
due to multiple acceleration, or an increase in the strength of the shocks.
Implications of our results for non-thermal X-ray emission are discussed.
}

\begin{document}
\maketitle

\section{Introduction}
Many hot stars of spectral type O and B are observable at radio wavelengths, 
due to thermal emission from the circumstellar ionized gas in the stellar wind. 
The thermal radiation is free-free emission (Bremsstrahlung) from an electron 
accelerated in the Coulomb field of an ion. The emergent radio spectrum has a
characteristic spectral index\footnote{The radio spectral index $\alpha$ is 
defined by $F_\nu \propto \nu^\alpha \propto \lambda^{-\alpha}$} $\alpha \approx 0.6$ 
(Wright~\&~Barlow~\cite{WB75}), which fits the observations for most O stars 
quite well.

However, 25\% of the brightest O stars have a radio spectrum with a spectral index
which is very different from the thermal wind emission (Abbott~et~al.~\cite{ABC84}). 
It is believed that this \emph{non-thermal} radio emission is synchrotron radiation 
by relativistic electrons (White~\cite{W85}). The preferred mechanism to accelerate 
electrons to relativistic energies is particle acceleration in shocks (Fermi~\cite{F49}), 
as shocks are known to exist in stellar winds. In single stars, shocks are generated 
by the instability of the driving mechanism of the wind (Owocki~\&~Rybicki~\cite{OR84}). 
For binaries, shocks also arise in colliding winds (Eichler~\&~Usov~\cite{EU93}). 
Non-thermal radio emission from colliding wind binaries has been further 
investigated by Dougherty~et~al.~(\cite{D03}). In this paper we limit ourselves 
to single stars.

Besides shocks, a magnetic field is needed to produce a synchrotron spectrum.
Magnetic fields of normal hot stars are below current detection limits 
($\sim 100~{\rm Gauss}$). It is plausible that a small magnetic field is present 
in the star,  either the fossil magnetic field, or a field created by a dynamo 
mechanism (MacGregor~\&~Cassinelli~\cite{MC03}).

The model proposed by White~(\cite{W85}) was further developed by
Chen~\&~White~(\cite{CW91},~\cite{CW94}), who showed that a synchrotron model can
successfully reproduce the radio spectrum of \object{9~Sgr} and \object{Cyg~OB2~No.~9}.

In the present paper, we investigate what properties a synchrotron model
should have in order to reproduce an observed non-thermal radio spectrum.
More specifically, we look at {\em all} models that fit a given set of
observations and thus derive the permitted range of a number of parameters.
The stringency of the constraint on the parameters depends critically on the
quality of the available observations, as well as on the number of observed
wavelengths. We selected  \object{Cyg~OB2~No.~9} as the object of this study because a
set of three simultaneous observations (at 2, 6 and 20 cm) with small
error bars is available.

For the sake of clarity it is desirable that the properties of the model be
contained in a limited number of parameters with a clear physical
interpretation. Rather than calculating the momentum distribution of
relativistic electrons from a model of the shocks in the stellar wind, we
use a power-law dependence on the compression ratio. The
uncertainty about the exact shock structure is so large that an ab initio
calculation of the momentum distribution would lead to a greater complexity of
the model without increasing its physical accuracy.

The remainder of this paper is organised as follows. In Sect.~\ref{sect:model} 
we describe the model for the synchrotron radiation in detail. The effect of 
cooling mechanisms on the momentum distribution is investigated in 
Sect.~\ref{sect:cooling}. In Sect.~\ref{sect:applications} we apply the model to
\object{Cyg~OB2~No.~9}.  Finally, we discuss the results in Sect.~\ref{sect:conclusions}
and give some conclusions.

\section{The model}\label{sect:model}
\subsection{Power}
In the presence of shocks and a magnetic field, electrons are accelerated to 
high velocities (Bell~\cite{B78}) through the first-order Fermi mechanism 
(Fermi~\cite{F49}) and emit synchrotron radiation as they gyrate around the 
magnetic field lines at relativistic velocities. The synchrotron power radiated 
per unit volume per unit frequency by a single electron with momentum $p$, 
moving in a magnetic field $B$ (dependent on distance $r$) with pitch angle 
$\theta$, can be expressed as
(Westfold~\cite{W59})
\begin{eqnarray}\label{eq:power}
        P_{\nu}(p,\theta,r)&=&\frac{\sqrt{3}e^3}{m_{\rm e}c^2}Bf(\nu,p) \sin\theta \nonumber\\
        &&\times \;\;\;\;\quad\quad F\left(\frac{\nu}{f^3(\nu,p)\nu_s(p,r)\sin\theta}\right),
\end{eqnarray}
where $m_{\rm e}$ and $e$ are the electron mass and charge, $c$ the speed of 
light, $\nu$ the frequency, $F(x)=x\int_{x}^{\infty}{d\eta K_{5/3}(\eta)}$ 
and $K_{5/3}$ the modified Bessel function. For the meaning of $f(\nu,p)$, 
see Sect.~\ref{sect:razin}. The synchrotron power spectrum extends up to 
frequencies of order $\nu_s$ before falling away, with the critical frequency 
$\nu_s$ defined as
\begin{equation}\label{eq:critfreq}
        \nu_s(p,r)=\frac{3}{4\pi}\frac{eB}{m_{\rm e}c}\left(\frac{p}{m_{\rm e}c}\right)^2.
\end{equation}

\subsection{Razin effect}\label{sect:razin}
The presence of $f(\nu,p)$ in the synchrotron power is a consequence of the 
Tsytovitch-Razin effect (or Razin effect for short; Tsytovitch~\cite{T51}; 
Razin~\cite{R60}). Relativistic beaming plays a dominant r\^ole in the 
explanation of synchrotron emission. In the presence of a cold plasma, the
refraction index is smaller than unity. This increases the beaming angle, 
thereby reducing the beaming effect. Thus the synchrotron emitting power is 
greatly reduced. Ginzburg~\&~Syrovatskii~(\cite{GS65}) expressed $f(\nu,p)$ as
\begin{equation}\label{eq:razin}
        f(\nu,p)=\left[1+\frac{\nu_0^2}{\nu^2}\left(\frac{p}{m_{\rm e}c}\right)^2\right]^{-1/2},
\end{equation}
with $\nu_0=\sqrt{\frac{n_{\rm e} e^2}{\pi m_{\rm e}}}$ the plasma frequency and $n_{\rm e}$ 
the number density of thermal electrons.

In the absence of the Razin effect ($f(\nu,p)=1$), the essential part of 
the synchrotron emission of an individual electron with momentum $p$ is 
emitted below the critical frequency $\nu_s$, with a maximum at
\begin{equation}\label{eq:stralingsgebied}
       \nu=0.29\nu_s \sin \theta \approx 1.3 \left(\frac{B\sin \theta}{1~{\rm Gauss}}\right) 
	\left(\frac{p}{m_{\rm e}c}\right)^2 {\rm MHz}.
\end{equation}
Away from the maximum, the synchrotron radiation falls off rapidly. We can 
therefore use Eq.~(\ref{eq:stralingsgebied}) to estimate the momentum range 
relevant for the radio spectrum. For a magnetic field of 0.1~Gauss, radiation 
between $2-20~{\rm cm}$ ($15-1.4~{\rm GHz}$) comes from electrons with 
momenta between $\approx 50-150~{\rm MeV/c}$, where we set $\sin \theta =1$.

From Eq.~(\ref{eq:razin}) we see that the Razin effect is important when 
$\nu\ll \nu_0 p/(m_{\rm e}c)$. If on the other hand $\nu \gg \nu_0 p/(m_{\rm e}c)$, 
the influence of the Razin effect is small. We can then use Eq.~(\ref{eq:critfreq}) 
to substitute $p$ and find the frequencies for which the influence of the 
Razin effect is small.  In the literature (e.g. Ginzburg~\&~Syrovatskii~\cite{GS65}), 
this is given by
\begin{equation}\label{eq:freqvoorwaarde}
        \nu \gg  20 \frac{n_{\rm e}}{B}
\end{equation}
where, in view of the approximate nature, the factor $0.29$ has not been taken into 
account. In a typical hot-star wind, the thermal electron density at $100~R_*$ 
is of order $10^7~{\rm cm^{-3}}$. For a magnetic field of $0.1~{\rm Gauss}$ 
this means that the Razin effect begins to suppress radiation at wavelengths 
larger than $15~{\rm cm}$, well within the observable range. For smaller magnetic 
fields, the radiation is suppressed at even smaller wavelengths e.g.  $1.5~{\rm cm}$ 
for a $0.01~{\rm Gauss}$ field.

\subsection{Flux}\label{sect:flux}
The synchrotron emissivity of a distribution $N(p,r)$ of relativistic electrons
 can be expressed as
\begin{equation}\label{eq:emissivity}
	j_\nu(r)=\frac{1}{4\pi}\int_{p_0}^{p_c}{\mathrm{d}p N(p,r) \bar{P_{\nu}}(p,r)},
\end{equation}
where $\bar{P_{\nu}}(p,r)$ is the synchrotron power $P_\nu(p,\theta,r)$ 
averaged over solid angle assuming that the electron velocity distribution is 
locally isotropic. The low (high) momentum cut-off of the electron distribution 
is defined by $p_0$ (respectively $p_c$). We will specify the cut-off values 
in Sect.~\ref{sect:cooling}.

The emergent flux is given by
\begin{equation} \label{eq:flux}
       F_\nu=\frac{1}{D^2}\int_{R_\nu}^{R_{\rm max}}{\mathrm{d}r 4\pi r^2 j_\nu(r)},
\end{equation}
where $D$ is the distance to the star. The integration boundaries reflect the 
fact that, at any given radio frequency, only a limited part of the wind contributes 
to the flux. The choice of these boundaries is rather subtle. The lower boundary
mimics the effect of the large free-free optical depth of the wind, which effectively
shields the inner part of the synchrotron emitting region from observation. 
The lower boundary must increase with increasing wavelength, to reflect the 
increasing free-free optical depth of the wind. It is customary
(e.g. White~\cite{W85}) to use the characteristic radius from 
Wright~\&~Barlow~(\cite{WB75}) as the lower boundary. However, this radius 
corresponds to a small optical depth ($\tau_{\nu} \approx 0.25$) and was introduced 
to produce the correct \emph{thermal} radio flux with a simple one-dimensional 
integral. Its physical meaning is often misstated and is actually quite limited. 
It is \emph{not} true that all observed radio emission originates from above $R_{\nu}$. 
It is quite easy to replace Eq.~(\ref{eq:flux}) by the correct expression which is, 
however, rather cumbersome. We therefore relegate its derivation to Appendix A and 
prefer to use the simple approximation Eq.~(\ref{eq:flux}) in the discussion of our 
results. The calculations were done using the exact expression given in 
Eq.~(\ref{eq:flux1d}).

As for the outer integration boundary, it is well known that instability-generated 
shocks decay as they move out with the wind (e.g. Runacres~\&~Owocki~\cite{RO02}). 
It can therefore be expected that they eventually become too weak to accelerate 
electrons to relativistic energies. Because relativistic electrons rapidly lose 
momentum as they move away from the shock (Sect.~\ref{sect:emittingregion}), the
outer boundary of the synchrotron emission region ($R_{\rm max}$)
can also be visualised as the position of the last shock strong enough to accelerate 
electrons to relativistic energies. It does not depend on frequency and is a 
fundamental parameter of our models.

The total emergent flux has a minor contribution from free-free emission by
\emph{thermal} electrons. This contribution is calculated using the Wright~\&~Barlow 
formalism (Wright~\&~Barlow~\cite{WB75}) and added to the non-thermal flux.

\subsection{Number distribution and magnetic field}\label{sect:distributionandfield}
In order to derive an explicit expression for the emissivity, we need to specify
the number distribution $N(p,r)$ and the magnetic field $B(r)$.

It was shown by Bell~(\cite{B78}) that the acceleration of electrons through the Fermi
mechanism results in a power-law momentum distribution with exponent $n$, where $n$ is
a function of the compression ratio $\chi$ of the shock: $n=(\chi+2)/(\chi-1)$.
For reasons of simplicity, we also assume that the distribution is a continuous
function of distance, given by a power law with exponent $\delta$. $N(p,r)$ is then 
expressed as
\begin{equation}\label{eq:distribution}
        N(p,r)=N_0 \left(\frac{r}{R_*}\right)^{-\delta} p^{-n},
\end{equation}
where $N_0$ is a normalisation constant related to the number of relativistic electrons.
The total number of relativistic electrons at the stellar surface is given by 
$\int_{p_0}^{p_c}{\mathrm{d}p N_0\, p^{-n}}$. This gives
\begin{equation}\label{eq:normalisation}
	N_0=f_* n_{\rm e}^* (n-1)\, p_0^{n-1},
\end{equation}
where we introduced $f_*$ and $n_{\rm e}^*$ as the fraction of relativistic electrons and 
the total electron number density at the stellar surface. We also assumed $p_c\gg p_0$
(see Sect.~\ref{sect:cooling}). The radial variation of the fraction of
relativistic electrons in the wind is described by $f_* (r/R_*)^{-\delta+2}$.
It must be emphasised that $f_*$ is merely a parameter used to set the overall 
level of this fraction. It should not be too strictly interpreted as a physical quantity. 
Indeed, close to the stellar surface,  the relativistic electron population may be 
strongly suppressed by inverse Compton cooling in the intense UV-radiation. Therefore 
$f_*$ need not represent the actual relativistic electron fraction at the stellar surface.

For $n_{\rm e}^*$ we use
\begin{equation}
	n_{\rm e}^*=\frac{\gamma \dot{M}}{4\pi R_*^2 v_\infty\mu m_{\rm H}},
\end{equation}
where $\dot{M}$ is the mass loss rate, $v_\infty$ the terminal velocity of the stellar wind,
$m_{\rm H}$ the mass of a hydrogen atom, $\gamma$ the mean number of electrons per ion
and $\mu$ the mean ionic weight. For simplicity we assume $\gamma=1$ and $\mu=1.4$.
Note that $n_{\rm e}^*$ is not the actual electron density at the stellar surface (as the
variation of the velocity has not been taken into account), but merely provides the 
correct scaling factor for the electron density at large distances.

In Eq.~(\ref{eq:power}) the magnetic field $B$ must be specified. Unfortunately,
the magnetic fields of hot stars are below current detection limits. Therefore, we can 
say little about their radial dependence. For want of anything better, we will
adopt the expression (Weber~\&~Davis~\cite{WD67})
\begin{equation}\label{eq:magnetic}
        B(r)=B_* \frac{v_{\rm rot}}{v_{\infty}}\frac{R_*}{r},
\end{equation}
where $B_*$ is the surface magnetic field and $v_{\rm rot}$ the rotational velocity 
of the star.
This expression is only valid at large
distances from the star ($r\geq10~R_*$), which covers the region where the 
synchrotron radiation is emitted. For the surface magnetic field, the detection 
limit, which is of order $\sim 100~{\rm Gauss}$ (Mathys~\cite{M99}), serves as an 
upper limit. A tentative lower limit can be derived from the synchrotron emission 
itself. At local magnetic fields smaller than $\sim 0.005~{\rm Gauss}$ at 
$100~R_*$, all radio emission would be suppressed by the Razin effect. Using 
Eq.~(\ref{eq:magnetic})  and $v_{\rm rot}/v_\infty=250/2900$, this translates 
to a surface magnetic field of order $5~{\rm Gauss}$.

\begin{figure}
\resizebox{\hsize}{!}{\includegraphics{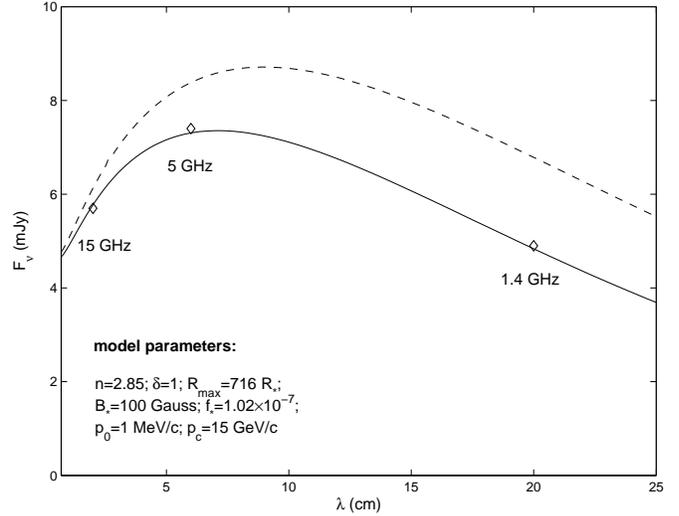}}
\caption{The full line represents a model that fits the VLA-observations
        (21~Dec~1984) for \object{Cyg~OB2~No.~9} (Bieging et al.~\cite{BAC89}), while
	the dashed line neglects the Razin effect. The diamonds are the observations 
	and the error bar is given by the size of the symbol. The stellar parameters
	are listed in Table~\ref{tab:CygOB2No9}.}
\label{fig:fluxexample}
\end{figure}

Once a choice for the parameters $B_*$, $n$, $\delta$, $R_{\rm max}$, $f_*$, $p_0$ 
and $p_c$ has been made, Eq.~(\ref{eq:flux1d}) can be integrated numerically. The 
total emergent flux (i.e. non-thermal + thermal) for one such model is shown in 
Fig.~\ref{fig:fluxexample}.  

\subsection{Tests}  \label{sect:tests}
If we neglect the Razin effect and approximate the cut-off values $p_0$ and 
$p_c$ by $0$ and $\infty$, the emissivity can be calculated analytically (e.g.
Rybicki~\&~Lightman~\cite{RL79}) from Eq.~(\ref{eq:emissivity}). The expression 
for the emissivity then becomes a power law: $j_\nu \propto \nu^{-(n-1)/2}$.
Introducing this into Eq.~(\ref{eq:flux}), an analytic expression is found for 
the synchrotron flux (Ginzburg~\&~Ozernoy~\cite{GO66}), which can then be compared 
to the results of the numerical code.

In the code, the momentum integral of the emissivity, given in Eq.~(\ref{eq:emissivity}), 
is computed using the trapezoidal rule and the step size is refined until it 
reaches a desired fractional precision of $10^{-3}$. This emissivity has to be 
calculated at different positions in the wind. Then, the spatial integral of the 
flux, Eq.~(\ref{eq:flux}), is treated in the same way as the momentum integral, but 
up to a precision of order $1\%$. The number distribution is given by 
Eq.~(\ref{eq:distribution}), while the mean synchrotron power $\bar{P_\nu}$ is 
calculated using a linear interpolation between tabulated values. This way, the
mean synchrotron power $\bar{P_\nu}$ needs to be calculated only once (as a function
$\nu/f^3(\nu,p)\nu_s(p,r)$) which reduces the calculation time significantly. The next 
step is to approximate the  momentum cut-off values, $p_0$ and $p_c$ by $0$ and 
$\infty$. We start with $p_0= 1~{\rm MeV/c}$ and $p_c=10~{\rm GeV/c}$ and then we 
expand the interval by doubling the value for $p_c$ and halving the value of $p_0$ 
until the solution for the integral has converged. We can then compare the numerical 
results to the analytic solution. We calculated $5000$ models with $B_*$ in the range 
10--100~Gauss, $n$ between 1.5--7.5, $\delta$ between 0--5 and $R_{\rm max}$ between 
100--10\,000~$R_*$. The relative errors of the numerical models compared to the 
analytic models did not exceed the 1\% level.

\section{Cooling mechanisms}\label{sect:cooling}
\subsection{Practical momentum limits}
In this section we determine practical integration limits for the momentum 
distribution, which we need to calculate the synchrotron emissivity, given 
by Eq.~(\ref{eq:emissivity}).

Relativistic electrons can lose their energy through a variety of cooling mechanisms.
Coulomb collisions, inverse Compton scattering, adiabatic cooling, synchrotron cooling
and Bremsstrahlung cooling can all play a r\^ole. The effect of these cooling mechanisms 
has been discussed in detail by Chen~(\cite{C92}). At the high energy end of the 
spectrum, electrons are mainly cooled by inverse Compton scattering. This sets an upper 
limit $p_c$ to the momentum range. At the low (non-relativistic) end of the spectrum, 
Coulomb collisions are the dominant cooling mechanism. This sets a lower boundary $p_0$ 
to the momentum range. As momenta below $p_0$ or above $p_c$ cannot occur, 
the momentum range is limited to $[p_0,p_c]$. Only a fraction of this range actually 
contributes to observable radio emission, as can be seen from 
Eq.~(\ref{eq:stralingsgebied}).
For $B_*=100~{\rm  Gauss}$ and assuming that the synchrotron radiation is formed somewhere 
between 30 and $1000~R_*$ (see Sect.~\ref{sect:rmax}), the range of momenta that contribute 
to the observable radio flux is $30-500~{\rm MeV/c}$. 
It is therefore convenient to introduce an ad-hoc integration interval $[p_1,p_2]$
that is contained in $[p_0,p_c]$ but is large enough to include all momenta
that produce observable radio emission. Within these constraints we choose 
$[p_1,p_2]$ as small as possible, for reasons of computational efficiency.
The exact choice of $p_1$ and $p_2$ is unimportant.
In the following section we shall derive values of $p_1$ and $p_2$ that can be 
used in the models.

\subsection{Low momentum cut-off}\label{sect:lowmomentumcutoff}
Synchrotron emission at wavelengths shorter than 20~cm is produced by relativistic 
electrons with momenta 
above $\approx~30~{\rm MeV/c}$. Any value $p_1\ll 30~{\rm MeV/c}$ can therefore be 
used as a practical lower limit to the momentum range. A convenient choice of $p_1$ is
$p_1=1~{\rm MeV/c}$. Although the choice of $p_1$ does not influence the flux,
it does influence the number of electrons (as seen in Eq.~(\ref{eq:normalisation})).

We still need to check whether electrons of momentum $p_1$ exist, i.e. whether 
$p_1>p_0$ (where $p_0$ is the momentum cut-off set by Coulomb cooling). 
We therefore compare the acceleration time of electrons with momentum 
$p_1=1~{\rm MeV/c}$ with the time-scale for Coulomb collisions of electrons with thermal ions. 
Collisions happen 
on time-scales given by (Spitzer~\cite{S56})
\begin{equation}\label{eq:defmfp}
        t_{\rm D}=\frac{m_{\rm e}^2 v^3}{8\pi e^4 n_{\rm g}\ln\Lambda},
\end{equation}
where $v$ is the speed of the relativistic electrons, $n_{\rm g}$ the number density 
of the ions and $\ln\Lambda$ the
Coulomb logarithm. In hot-star winds the Coulomb logarithm is $\sim 10$. 
The time to accelerate an electron up to a momentum $p$ is given by
(Chen~\cite{C92})
\begin{equation}
     	t_{\rm acc}=\frac{3}{2\Delta u(r)}\frac{pc}{eB},
\end{equation}
where $\Delta u(r)$ is the shock velocity difference at radius $r$. 
For an electron of 1~MeV/c at a radial distance of $100~R_*$ where the number density 
of the ions is $n_{\rm g}=10^7~{\rm cm^{-3}}$ 
and $B=0.1$~Gauss, we find $t_{\rm D}\approx 10^5~{\rm s}$, while 
$t_{\rm acc}\approx 5\times10^{-3}~{\rm s}$ (where we assume
$\Delta u=100~{\rm km~s^{-1}}$). Further out in the wind the ratio $t_{\rm D}/t_{\rm acc}$
increases. This means that electrons with momentum 1~MeV/c will not be Coulomb-cooled 
significantly.

\subsection{High momentum cut-off}\label{sect:highmomentumcutoff}
Synchrotron emission at wavelengths greater than 2~cm is produced by 
relativistic electrons with momenta below
$\approx~500~{\rm MeV/c}$ (for $B_*=100~{\rm Gauss}$). Any value 
$p_2 \gg 500~{\rm MeV/c}$ can therefore be used as a practical upper limit to the 
momentum range, providing $p_2<p_c$ (where $p_c$ is the high momentum cut-off).

The highest momentum attainable for relativistic electrons is determined by the 
balance between acceleration and inverse Compton scattering at the shock. 
This results in a maximum attainable momentum $p_c$, that can be expressed as (Chen~\cite{C92})
\begin{equation}\label{eq:Chencutoff}
	p_c(r)= m_ec\left(4\pi\frac{eB(r)}{\sigma_{\rm T}L_*}\Delta u(r) r^2\right)^{1/2},
\end{equation}
where $\sigma_{\rm T}$ is the Thomson cross section and $L_*$ the stellar bolometric luminosity.
Using typical hot star values
($L_*=10^6~{\rm L_{\sun}}$, $v_\infty=2\,500~{\rm km~s^{-1}}$, $v_{\rm rot}=200~{\rm km~s^{-1}}$, 
$R_*=20~{\rm R_{\sun}}$ and 
$\Delta u(r)=100~{\rm km~s^{-1}}$), we can estimate a numerical value for the high momentum cut-off,
\begin{equation}\label{eq:highcutoff}
	p_c(r)\approx 0.97 \frac{\rm GeV}{c} \left(B_*\frac{r}{R_*}\right)^{1/2}.
\end{equation}
For $B_*=100~{\rm Gauss}$ and $r=10~R_*$, Eq.~(\ref{eq:highcutoff}) gives $p_c=30~{\rm GeV/c}$.
At larger distances from the star $p_c$ is larger. A practical choice for $p_2$ that 
satisfies $p_2<p_c$ is $p_2=15~{\rm GeV/c}$. 

Now that a choice has been made for $p_1$ and $p_2$, the independent parameters are
reduced to $B_*$, $n$, $\delta$, $R_{\rm max}$ and $f_*$.

\subsection{Emitting region}\label{sect:emittingregion}
We have estimated practical upper and lower limits to the momentum range that are 
compatible with the high and low momentum cut-offs due to Compton and Coulomb cooling, 
respectively. Of course, cooling can also change the distribution between these limits. 
Specifically, as the relativistic electrons move away from the shock in which they were 
accelerated, inverse Compton scattering will rapidly cool the most energetic electrons. 
Inverse Compton scattering does not change the total number of relativistic particles, but
quickly reduces their energy below the energy of observable radio emission. Thus,
the synchrotron emitting gas is concentrated in more or less narrow regions behind the
shocks. For the sake of simplicity we neglect this layered structure of the synchrotron 
emitting region and assume a continuous volume of relativistic particles. The adequacy of this 
assumption is discussed in Sect.~\ref{sect:conclusions}.

\section{Application and results}\label{sect:applications}
\begin{table}[t]
 \caption{Relevant stellar parameters for \object{Cyg~OB2~No.~9} adopted in this paper. The numbers
	are taken from Bieging~et~al.~(\cite{BAC89}, BAC), Leitherer~(\cite{L88}, L) and
	Herrero~et~al.~(\cite{HCVM99}, HCVM).}
\label{tab:CygOB2No9}
\begin{center}
\begin{tabular}{lrl}
	\hline
	\hline
       	Parameter & &Ref.\\

   	\hline
	$v_{\rm rot} \sin i~({\rm km~s^{-1}})^a$&145&BAC\\
	$T_{\rm eff}~({\rm K})^b$&44\ 500&HCVM\\
        $R_*~({\rm R_{\sun}})$&	22&HCVM\\
	$d~({\rm kpc})$&1.82&BAC\\
	$v_\infty~({\rm km~s^{-1}})$&2\ 900&	L\\
	$\dot{M}~({\rm M_{\sun}~yr^{-1}})$&$2.0\times 10^{-5}$&L\\
	\hline
\end{tabular}
\end{center}

$^a$ We follow White~(\cite{W85}) in assuming that $v_{\rm rot}= 250~{\rm km~s^{-1}}$, which
is not in contradiction with the value for $v_{\rm rot} \sin i$. \\
$^b$ The wind temperature is assumed to be $0.3$ times the stellar effective temperature
(Drew~\cite{D89}). \\
 \end{table}

\begin{table}[t]
 \caption{With the stellar parameters for \object{Cyg~OB2~No.~9} given in Table~\ref{tab:CygOB2No9},
 we have the following values for the characteristic radius $R_\nu$ (see Eq.~(\ref{eq:charradius})) and
  for the free-free emission from the wind (Wright~\&~Barlow~\cite{WB75}). For 
simplicity we assume $\gamma=1, Z^2=1$ and $\mu=1.4$. As a reference we also give
the values of the non-thermal component to the flux and the observed flux with error bars.}
\label{tab:charradius}
\begin{center}
\begin{tabular}{ccccccc}
   \hline
   \hline
   $\lambda$&$\nu$&  &$R_\nu(T_{\rm wind})$& $F_\nu^{\rm ff}$& $F_\nu^{\rm nt}$& $F_\nu^{\rm obs}$\\
   ({\rm cm})&(GHz)& & ($R_*$)&        ({\rm mJy})&      ({\rm mJy})     &  ({\rm mJy})\\
   \hline
   2&15&        &124                    &1.1            &4.6              &5.7 $\pm$0.1\\
   6&5&         &266                    &0.6            &6.8              &7.4 $\pm$0.1\\
   20&1.4&      &614                    &0.3            &4.6              &4.9 $\pm$0.1\\
   \hline
\end{tabular}
\end{center}
\end{table}

\subsection{\object{Cyg~OB2~No.~9}: 21~Dec~1984 data}
We can now apply the model to \object{Cyg~OB2~No.~9} (O5 If). It was first discovered by 
Abbott~et~al.~(\cite{ABC84}) that the spectral index of its radio emission does not follow the
$\alpha = 0.6$ law expected for a thermal wind source (see Fig.~\ref{fig:fluxexample}).
Also, the radio emission of \object{Cyg~OB2~No.~9} is highly variable. On one occasion the radio 
spectrum even showed a typical free-free spectrum (with $\alpha = 0.6$). If we determine 
the mass loss rate from this radio observation, we find the same value as derived from 
H$\alpha$ (Leitherer~\cite{L88}). So we can assume that at that time we saw the underlying 
free-free emission from the stellar wind. Of the observational radio data that are available for
\object{Cyg~OB2~No.~9}, we will use the observation that has detections of emission in three radio 
wavelength bands. We chose the 21 Dec 1984 observation with the VLA (Bieging~et~al.~\cite{BAC89}) 
because of its small error bars on the detections. By fitting these observations we can 
derive the solution space of the model parameters. The stellar parameters adopted in our 
model are listed in Table~\ref{tab:CygOB2No9}.

\subsection{Results}
We recall that the parameter space is spanned by $n$, $\delta$, $f_*$, $R_{\rm max}$ and $B_*$.
The parameters have a straightforward interpretation: $n$ corresponds to a typical compression 
ratio in the synchrotron formation region, $\delta$ describes the spatial dependence of the number 
distribution, $f_*$ describes the fraction of relativistic electrons at the surface and 
$R_{\rm max}$ is the outer boundary of the synchrotron emission region. We take 
$B_*=100~{\rm Gauss}$ in the fitting procedure, which corresponds to a local magnetic 
field of $B=1.7\times 10^{-2}~{\rm Gauss}$ at $r=500~R_*$.

\begin{figure}
\resizebox{\hsize}{!}{\includegraphics{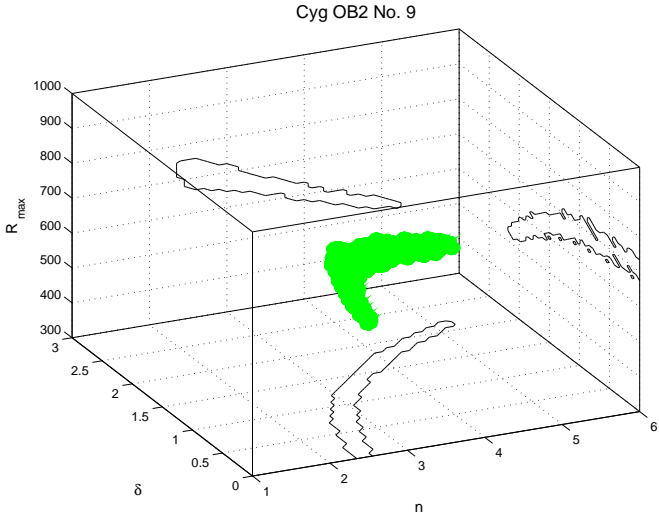}}
\caption{All the combinations of $n$, $\delta$ and $R_{\rm max}$ with $B_*= 100~{\rm Gauss}$ 
	that fit the VLA-observations (21 Dec 1984) for \object{Cyg~OB2~No.~9} lie within the 
	boomerang-shaped region. Projections on different planes are plotted to situate 
	the solutions in the parameter space. }
\label{fig:VLA21Dec1984}
\end{figure}

We calculate the synchrotron flux in the three wavelengths for a grid in $n$, $\delta$, 
$R_{\rm max}$ and $f_*$ and select those combinations of the parameters that fit the 
VLA-observations within the error bars. This procedure produces a strong constraint 
on $n$, $\delta$ and $R_{\rm max}$, but not on $f_*$. The parameter $f_*$ is well constrained
locally (i.e. for each combination of $n$, $\delta$ and $R_{\rm max}$), but not globally 
(i.e. for the entire range of $n$, $\delta$ and $R_{\rm max}$). This parameter does play 
a minor r\^ole in limiting the other parameters as the number of relativistic electrons 
cannot exceed the total number of electrons. In principle this limitation depends on
$p_1$, but in practice the effect is minimal. All possible combinations of $n$, $\delta$ 
and $R_{\rm max}$ lie within the boomerang-shaped region in Fig.~\ref{fig:VLA21Dec1984}.

\subsubsection{$R_{\rm max}$}\label{sect:rmax}
The most important result from Fig.~\ref{fig:VLA21Dec1984} is that \emph{$R_{\rm max}$ is 
well constrained} and almost not influenced by $n$ and $\delta$. The outer boundary of 
the synchrotron emitting region must lie between $520~R_*$ and $750~R_*$.
To understand this result, it is important to realise that the different wavelengths
play a different r\^ole in the fitting procedure. The fluxes at 2 and 6~cm constrain
$n$, $\delta$ and $f_*$. The parameters $n$ and $\delta$ determine the slope between
2 and 6~cm and $f_*$ sets the global flux level. Note that different combinations of $n$ 
and $\delta$ can give the same slope. The effect of $R_{\rm max}$ on the slope is negligible. 
The flux at 20~cm is fitted by the only remaining parameter $R_{\rm max}$. The strong 
dependence of the 20~cm-flux on $R_{\rm max}$ is due to the larger free-free opacity at 
larger wavelengths. This means that more of the synchrotron emitting region is shielded 
at 20~cm than at 2 and 6~cm. This also explains why $R_{\rm max}$ has only a small influence 
on the 2 and 6~cm-fluxes. Note that the range in $R_{\rm max}$ (520 -- 750~$R_*$) 
extends below the characteristic radio radius at 20~cm ($R_\nu(20~{\rm cm})=614~R_*$). 
This is not a contradiction as radio emission from  below $R_\nu$ can be detected 
(see Sect.~\ref{sect:flux}).

\begin{figure}
\resizebox{\hsize}{!}{\includegraphics{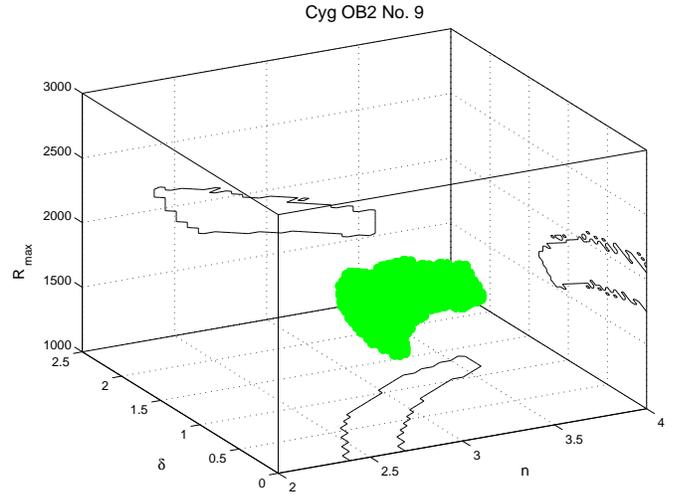}}
\caption{Similar figure as Fig.~\ref{fig:VLA21Dec1984}, but now calculated for a surface
        magnetic field $B_*=10~{\rm Gauss}$ (Note the different scales compared to 
	Fig.~\ref{fig:VLA21Dec1984}). }
\label{fig:VLA21Dec1984b}
\end{figure}

While $R_{\rm max}$ is insensitive to $n$ and $\delta$, it depends strongly on the magnetic 
field. Assuming a smaller surface magnetic field of 10~Gauss rather than the 100~Gauss used
 previously, we find a larger value of $R_{\rm max}$ (see Fig.~\ref{fig:VLA21Dec1984b}). 
$R_{\rm max}$ is still well constrained, but now the range extends from $1\,500~R_*$ 
up to $2\,200~R_*$. The larger value of $R_{\rm max}$ is caused by the greater importance 
of the Razin effect at lower magnetic fields. The 20~cm-flux is already greatly reduced by 
the Razin effect. To replenish the photons taken away by the Razin effect, the synchrotron 
emitting region must be larger than the value of $R_{\rm max}$ found for a magnetic field of
100~Gauss. It is not possible to fit the observations with a surface magnetic field of 
5~Gauss, because essentially all radio emission is suppressed by the Razin effect 
(Sect.~\ref{sect:distributionandfield}). In this sense the value of $B_*=10~{\rm Gauss}$ 
can be considered a lower limit for the surface magnetic field, and likewise the value 
found for $R_{\rm max}$ is an upper limit to the outer boundary of the synchrotron emission region. 

\subsubsection{$n$ and $\delta$}
Fig.~\ref{fig:VLA21Dec1984} shows solutions with $n$ extending from 2.5 to 5. Using 
the relation $\chi=(n+2)/(n-1)$, this translates to compression ratios from $\chi=3$ 
(moderately strong shocks) to $\chi=1.75$ (weak shocks).  Note that strong shocks 
($\chi=4$) cannot explain the observations. This is consistent with what was found by
Rauw~et~al.~(\cite{RB02}), who applied a similar model to \object{9~Sgr}. Since time-dependent hydrodynamical 
simulations (e.g. Runacres~\&~Owocki~\cite{RO02}) show a variety of shocks (and compression 
ratios) in the stellar wind, $\chi$ is some kind of compression ratio average. Stronger 
shocks however produce more radio emission than weaker shocks, because they accelerate 
more of the electrons into the momentum range where synchrotron radiation is emitted
at radio wavelengths. Thus the synchrotron spectrum is dominated by the stronger shocks.
Thus the value of $\chi$ we obtain should be interpreted as a maximum compression ratio
in the wind, rather than a mean compression ratio.

Note that most solutions have $\delta$-values lower than 2. This means that the fraction
of relativistic electrons increases further in the wind. This is surprising as one would 
expect a constant efficiency of the acceleration mechanism or even a decreasing efficiency
(due to the decay of the shocks in the wind). The increasing relativistic fraction points to
an increasing acceleration efficiency.  A possible explanation is multiple acceleration
of the electrons by subsequent shocks. Thermal electrons are injected in the shock front and
accelerated, but electrons, already accelerated by a previous shock, are also re-accelerated.
This would increase the number of relativistic electrons toward the outer region of
the stellar wind.

Another possible explanation is that, as hydrodynamical simulations show a variety of 
shocks, some stronger shocks might exist in the outer parts of the wind. This would 
increase the synchrotron emission which would show up as a $\delta<2$ in the parametrisation 
of the current model.

\subsubsection{Comparison with the Chen~\&~White model}
We can compare our results with the results found in Chen~\&~White~(\cite{CW91},~\cite{CW94}). 
We looked at all the models that fit the observations of \object{Cyg~OB2~No.~9} and derived the 
permitted range of the model parameters. We used a power law to describe the number 
distribution of relativistic electrons. Chen~\&~White calculated the number 
distribution from a model of shocks in the stellar wind. The parameters of their synchrotron 
model which reproduce the radio spectrum of \object{Cyg~OB2~No.~9} were $n=2.6$, $\delta<2$ and 
$R_{\rm max}=500R_*$. Even though we used different methods to describe the momentum 
distribution of relativistic electrons in the wind, the parameters of their synchrotron model 
are very similar and can be found near the solution space of our model.

\section{Discussion and conclusions}\label{sect:conclusions}
In this paper we have presented a theoretical model for non-thermal radio emission from single
hot stars, in terms of synchrotron emission from electrons accelerated in wind-embedded shocks. 
Applied to a specific observation of \object{Cyg~OB2~No.~9}, the model produces constraints on almost 
all of its parameters ($f_*$ being the only exception). A major result is that the outer 
boundary of the synchrotron emitting region ($R_{\rm max}$) is well constrained (520 -- 750~$R_*$ 
for $B_*=100$ Gauss). As the efficiency of Compton cooling prevents relativistic electrons from 
carrying their kinetic energy far from the shock, this means that our models predict that
shocks persist to at least $500~R_*$. This is consistent with recent hydrodynamical calculations 
(Runacres~\&~Owocki~\cite{RO03}).

Another interesting result is the absence of very strong shocks (those with compression ratio 
$\chi$ near 4) in the outer wind. We find shocks with $\chi = 3$ at most. This is consistent
with the hydrodynamical prediction that shocks become weaker as they move out with the stellar wind.

Within the range of solutions we find both $\delta \approx 2$ and $\delta < 2$. The physical 
picture associated with these different values of $\delta$ is rather different too. The
$\delta\approx 2$ solutions point to a constant fraction of relativistic electrons. It can be 
seen from Fig.~\ref{fig:VLA21Dec1984} that they also correspond to weak shocks ($\chi \la 2$). 
This offers a natural explanation for $R_{\rm max}$: the shocks weaken as they move
out with the flow and beyond $R_{\rm max}$ they are to weak to produce significant
synchrotron emission.

However, most of the solutions have $\delta < 2$, which points to the relativistic electron 
fraction increasing outwards. This could indicate an increasing compression ratio of the shocks, 
or at least the presence of a stronger shock in the outer part of the synchrotron emitting region. 
While this could incidentally have been the case at the time of observation, it does not sit well 
with the hydrodynamical prediction that shocks decay as they move outward. It is also hard to 
imagine why these shocks would rather suddenly disappear when they reach $R_{\rm max}$. A more 
plausible explanation of $\delta < 2$ is multiple acceleration: particles in the outer wind have 
had greater opportunity to pass through multiple shocks than particles in the inner wind. The 
$\delta < 2$ solutions are also those with larger compression ratios. In the case of multiple 
acceleration the momentum distribution is no longer a pure power law $\sim p^{-n}$, but a flatter 
function (White~\cite{W85}).  Simply put, this means that the momentum distribution for multiple 
acceleration by a number of weaker shocks resembles a distribution corresponding to a single 
stronger shock. This is then again consistent with the idea of an outer boundary of the synchrotron
emitting region, as all the shocks involved in the acceleration are weak.

In this paper we have assumed a power-law momentum distribution with a single compression ratio, 
where all of the volume can contribute to the emission. The efficiency of inverse-Compton 
cooling, however, will limit the synchrotron emission region to more or less narrow layers behind 
the shocks. The primary effect of this is to reduce the number of electrons in the momentum range 
relevant for radio emission. This means that the values of $f_*$ we find in our fitting procedure 
are underestimated as it is this parameter that sets the overall level of synchrotron emission, 
and must mimic the quenching of electrons by inverse Compton scattering. Because $f_*$ is not well
constrained by the adopted fitting procedure, this is of no consequence for our results. As a 
secondary effect, inverse-Compton cooling can also change the momentum distribution, as it 
affects the high momentum electrons more than the low momentum electrons. 
The importance of this effect will be investigated in a subsequent paper.

Hydrodynamical models predict a variety of compression ratios. Due to the strong dependence of 
the emission on the compression ratio, the strongest shock will dominate. In that case, the radiation 
would not come from the whole volume, but only from behind the strongest shocks. This may also 
explain the variability that is a characteristic feature of non-thermal radio emission 
(Bieging~et~al.~\cite{BAC89}). As the stronger shocks move in and out of the synchrotron emitting region
they lead to variability. 

The presence of relativistic electrons in the wind will also lead to non-thermal X-rays, due to 
the inverse Compton process (Chen~\&~White~\cite{CW91}). The non-thermal X-rays should show up 
as a hard X-ray tail in a spectrum. Clear evidence for such a non-thermal tail has remained elusive. 
Rauw~et~al.~(\cite{RB02}) show that the hard X-ray tail of the non-thermal radio emitter \object{9~Sgr} does not 
have the expected power-law dependence for non-thermal X-ray emission. The tail can also be modelled 
with a high-temperature ($\ge 2 \times 10^7$ K) \emph{thermal} plasma. 
It should be noted 
that the non-thermal X-ray emission is formed much closer to the stellar surface (where there is an 
abundant supply of UV photons) than the non-thermal radio emission. One could therefore surmise
that the shocks closer to the star are weaker and therefore result in an insufficient number of 
relativistic electrons to create detectable non-thermal X-rays. This suggestion, however, does not 
appear to be consistent with current hydrodynamical models, as the shocks are strongest in the inner 
wind and decay as they move outward. The hard X-ray tail of \object{9~Sgr} could also be explained by colliding 
wind emission from a long-period binary companion (Rauw~et~al.~\cite{RB02}).

To obtain non-thermal radio emission, all that is required is a magnetic field and shocks in 
the wind. As both should be present in any early-type star, the question arises why not all 
these stars are non-thermal emitters. One possible answer is that the magnetic field in 
thermal stars is lower. A lower magnetic field means that less synchrotron radiation is
generated and that the Razin effect is more efficient at suppressing it. This explanation 
suggests that searches to detect magnetic fields in O-type stars might have a better chance 
in non-thermal radio emitters. Another possible explanation is that the synchrotron emitting region
(essentially determined by the characteristic free-free radius and $R_{\rm max}$) is too small, 
so that the synchrotron radiation emitted falls below detection levels.

Finally, one needs to ask the uncomfortable question whether there is any such thing as 
synchrotron emission from \emph{single} O stars. For Wolf-Rayet stars, the connection between 
non-thermal emission and binarity is firmly established (Dougherty~\&~Williams~\cite{DW00}). 
For O stars, the situation is less clear. Roughly half of the non-thermal O stars are not 
known to be members of a binary system. However, the truism that absence of evidence is not 
evidence of absence obviously holds.  Also, there is an indication that one of the archetypal 
single non-thermal O star, \object{9~Sgr}, might actually be in a binary (Rauw~et~al.~\cite{RB02}). 
The possibility that all non-thermal emitting O stars are binaries cannot be excluded.

\begin{acknowledgements}
We thank S. Dougherty and J. Pittard for careful reading of the manuscript and S. Owocki for 
interesting discussions. SVL gratefully acknowledges a doctoral research grant by the 
Belgian State, Federal Office for Scientific, Technical and Cultural Affairs (OSTC). Part 
of this research was carried out in the framework of the project IUAP P5/36 financed by the OSTC.
\end{acknowledgements}

\appendix
\section{Flux: exact and approximated integral}\label{app:1}
We calculate the flux following Wright~\&~Barlow~(\cite{WB75}) for the free-free emission, but 
we replace $\chi B_\nu$, where $\chi$ is the free-free absorption coefficient and $B_\nu$ the 
source function (e.g. the Planck function), by the synchrotron emissivity $j_\nu$. The exact 
integral for the flux is
\begin{equation}\label{eq:flux2d}
       F_\nu=\frac{1}{D^2}\int_{0}^{R_{\rm max}}{\mathrm{d}q 2\pi q\int_{-\infty}^{\infty}{\mathrm{d}l
        j_\nu[r(l)] \mathrm{e}^{-\tau_\nu(q,l)}}},
\end{equation}
where $D$ is the distance to the star, $q$ the impact parameter, $l$ the distance along
the line of sight with an observer at $l=-\infty$ and $R_{\rm max}$ the outer boundary for
the synchrotron emission region. The optical depth $\tau_\nu(q,l)$ is defined as
\begin{equation}
      \tau_\nu(q,l)=\int_{-\infty}^{l}{\mathrm{d}l' \frac{K_{\rm ff}\gamma A^2}{(q^2+l'^2)^2}},
\end{equation}
with $A=\dot{M}/(4\pi\mu m_Hv_\infty)$, where the symbols have their usual meaning, and $K_{\rm ff}$ 
the free-free absorption coefficient given by Allen~(\cite{A73}). Note that by using $0$ as the 
lower integral boundary in Eq.~(\ref{eq:flux2d}), we in fact replaced the star by stellar wind 
material. This mathematical simplification has no consequence as the free-free optical depth is 
so large that any contribution of the star can be neglected anyway.

\begin{figure}
\resizebox{\hsize}{!}{\includegraphics{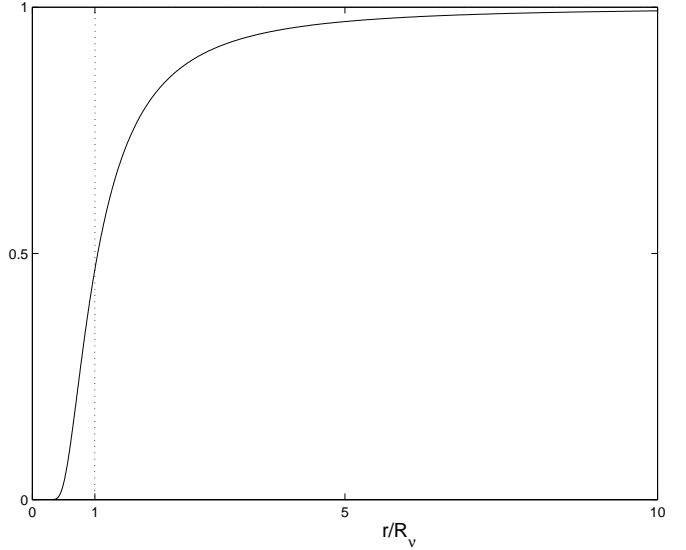}}
\caption{The full line is the geometric function $G(r/R_\nu)$, while the
	dashed line represents the Heaviside function.}
\label{fig:G}
\end{figure}

The equation for $\tau_\nu(q,l)$ can be integrated to (using Abramowitz~\&~Stegun~\cite{AS},~3.3.24)
\begin{equation}
	\tau_\nu(q,l)=\frac{K_{\rm ff}\gamma A^2}{2q^3}
	\left[\frac{l/q}{1+\frac{l^2}{q^2}}+\arctan\left(\frac{l}{q}\right)+\frac{\pi}{2}\right].
\end{equation}
Introducing polar coordinates,
\begin{eqnarray}\nonumber
	q&=&r\sin\theta \\
	l&=&r\cos\theta \\ \nonumber
	{\rm d}q{\rm d}l&=&r{\rm d}r{\rm d}\theta
\end{eqnarray}
we find
\begin{equation}
	\tau_\nu(q,l)=\frac{K_{\rm ff}\gamma A^2}{2r^3\sin^3\theta}(\sin\theta\cos\theta
	-\theta+\pi)
\end{equation}
and for the flux
\begin{equation}\label{eq:flux1d}
	F_\nu=\frac{1}{D^2}\int_{0}^{R_{\rm max}}{\mathrm{d}r4\pi r^2j_\nu(r)
	G\left(\frac{r}{R_\nu}\right)},
\end{equation}
with $R_\nu$ the characteristic radius of emission (Wright~\&~Barlow~\cite{WB75}) expressed as
\begin{equation}\label{eq:charradius}
        R_\nu=\frac{4}{\Gamma \left(\frac{1}{3}\right)\left(\frac{\pi}{2}\right)^{2/3}}(K_{\rm ff}\gamma A^2)^{1/3},
\end{equation}
and
\begin{eqnarray}
	G(x)&=&\frac{1}{2}\int_{0}^{\pi}{}\mathrm{d}\theta \sin\theta\\ \nonumber
	 &\times& \exp\left[-\frac{\left(\Gamma \left(\frac{1}{3}\right)/8\right)^3\pi^2}{x^3\sin^3\theta}
	(\sin\theta\cos\theta-\theta+\pi)\right].
\end{eqnarray}
The $\theta$ integration is only from $0$ to $\pi$, because the original integral only 
considered positive $q$ values (a factor of $2$ was included in the original integration 
to compensate for this). The function $G(r/R_\nu)$ is plotted on Fig.~\ref{fig:G}. Note 
that $G(r/R_\nu)$ can be approximated by a Heaviside function: $H=0$ if $r<R_\nu$ and 
$H=1$ if $r>R_\nu$. This brings us back to the approximation Eq.~(\ref{eq:flux}) for the flux.
The function $G(r/R_\nu)$ needs to be calculated only once, so that evaluating Eq.~(\ref{eq:flux1d})
does not take significantly more time than Eq.~(\ref{eq:flux}).

 \end{document}